\documentclass[showpacs,preprint]{revtex4-1}

\usepackage{amsmath}
\usepackage{amsfonts}
\usepackage{amssymb}

\newcommand{\ket}[1]{| #1 \rangle}
\newcommand{\bra}[1]{\langle #1 |}
\newcommand{\scalar}[2]{\left\langle #1 | #2 \right\rangle}
\newcommand{\vect}[1]{\mathbf{#1}}

\begin{document}

\title{Localizability in de Sitter space}

\author{N.\ Yokomizo}

\email{yokomizo@fma.if.usp.br}

\author{J.\ C.\ A.\ Barata}

\affiliation{Instituto de F\'{i}sica, Universidade de S\~{a}o Paulo, C.P.\ 66318, 05315-970 S\~{a}o Paulo-SP, Brazil}

\date{\today}


\begin{abstract}

An analogue of the Newton-Wigner position operator is defined for a massive neutral scalar field in de Sitter space. The one-particle subspace of the theory, consisting of positive-energy solutions of the Klein-Gordon equation selected by the Hadamard condition, is identified with an irreducible representation of de Sitter group. Postulates of localizability analogous to those written by Wightman for fields in Minkowski space are formulated on it, and a unique solution is shown to exist. Representations in both the principal and the complementary series are considered. A simple expression for the time-evolution of the operator Newton-Wigner is presented.

\end{abstract}

\maketitle

\section{Introduction}

The question of the existence and usefulness of a notion of localization for quantum particles moving at relativistic speed has a long history \cite{newton-wigner, wightman, philips,kaiser}. Although the idea of a position measurement is one of the most intuitive ideas of a quantum observable, there is no obvious mathematical counterpart to it in the relativistic domain. The conflict with intuition in such a fundamental subject is the main motivation to work on this problem. But there are also technical reasons for that. A quantum field theory is usually applied to the study of particle collision processes, so one needs to understand how to interpret the theory in terms of particles. The main question is how to assign probabilities for the detection of the produced particles in detectors placed at specific regions of space \cite{haag-swieca,buchholz,schroer,araki}, what amounts to defining a position probability distribution. Besides that, there is the very fact that classical particles do exist, i.e., that a classical limit of the underlying quantum theory exists which describes particles. A position operator is the natural tool to deal with this limit \cite{hepp,mcl}. Now the current widespread interest in quantum effects in curved spacetimes, boosted by experimental and theoretical discoveries in cosmology, motivates the analysis of the problem of localizability in a more general context. In particular, the present accelerated expansion of the universe \cite{riess,perlmutter} and the existence of an inflationary epoch in the very early universe \cite{weinberg,liddle,linde} suggest that there should be eras in the beginning of the universe and in the distant future when the geometry of the universe is approximately a patch of de Sitter space, what justifies our interest in this special geometry. Local effects of de Sitter geometry on particle dynamics have been investigated at the classical and quantum level \cite{moschella,bros,aldrovandi}.

In flat Minkowski space, an early solution to the problem was provided by the work of Newton and Wigner \cite{newton-wigner}, later reformulated in more rigorous terms by Wightman \cite{wightman}. It was proved that a natural set of postulates defines a unique position operator, at least for massive fields. But the operator found is frame-dependent, and alternative covariant notions of localizability were put forward since then \cite{philips, kaiser}. The interpretation of these operators and the possibility of actually measuring them have been discussed in the context of Quantum Field Theory in terms of specific models of interaction between a detector and the quantum field (see \cite{unruh-wald,marolf}). But if the particle moves in a curved spacetime, little is known. There are additional complications in the analysis, mainly due to the existence of multiple vacua. In fact, the concept of particle is not strictly necessary for Quantum Field Theory in Curved Spacetimes---the general theory can be formulated without introducing the notion of particles \cite{wald}. Only in special circumstances it still makes sense to speak of particles. In a flat Minkowski spacetime, for instance, that is certainly true. In the case of ultrastatic spacetimes, a notion of Newton-Wigner localization is available, as discussed in \cite{fulling}. And it is also natural that in regions where the curvature is small one should be able to speak of particle states---high-energy experiments are actually performed in a slightly curved space, and particles are observed. However, there is no clear specification of the necessary conditions for a particle interpretation to be available.

We have studied the case of a neutral massive scalar field in 2d de Sitter space, and have showed that a particle interpretation of this theory is possible. This problem was previously investigated exploring special decompositions of de Sitter group \cite{hannabuss} or an analogy with the Minkowski space case \cite{philips-wigner}. We have considered it in the context of the modern formulation of Quantum Field Theory in Curved Spacetimes, as described in \cite{wald,fulling}, for instance. Our strategy was the following. The main difficulty for the definition of particle states in a general curved spacetime lies on the existence of several inequivalent Fock representations for the canonical commutation relations. Since there is no preferred Fock representation, the concept of a particle becomes ambiguous. In de Sitter space, however, it is possible (see \cite{allen}) to select a unique vacuum state---the Bunch-Davies vacuum---by requiring: (i) physical states to satisfy the Hadamard condition, which corresponds to the requirement that the averaged energy-stress tensor can be renormalized by a point-splitting prescription \cite{wald}; and (ii) the vacuum to be invariant under the action of de Sitter group \cite{allen,BFH}. In this sense, there is a preferred Fock representation for massive free fields in de Sitter space. It is clear that the maximal symmetry of the de Sitter space is crucial for that.

There is an important remark that we should add here. In the Minkowski case, the notion of Newton-Wigner localization belongs to the realm of Relativistic Quantum Mechanics, not exactly to Quantum Field Theory, and is implemented in a one-particle Hilbert space. In curved space-times, however, the choice of the adequate one-particle Hilbert space, even for free fields, is dictated by the interest of the quantum field theoretical model one wishes to implement. Hence, the possibility to define a reasonable localization concept is also related to the choice of representations for the canonical commutation relations and, therefore, is not an exclusively relativistic quantum mechanical question.

We discuss the canonical quantization of the scalar field in the Bunch--Davies representation in Section II. An important fact is that the one-particle subspace $\mathcal{H}$ of the Fock representation can be interpreted as an irreducible representation of de Sitter group, as a result of vacuum invariance. That allowed us to write localizability conditions on $\mathcal{H}$ analogous to those formulated on irreducible representations of the Poincar\'{e} group in \cite{newton-wigner,wightman}. We describe these conditions in Section III, and prove that a unique solution exists at $t=0$. The position operator which satisfies these conditions is the natural analogue in de Sitter space of the Newton-Wigner position operator. These results are then compared with previous investigations of particle localization in de Sitter spacetime \cite{philips-wigner}, and the time-evolution of the position operator is described, based on an analogy with the case of Minkowski spacetime. Perspectives on future works are discussed in Section IV.

\section{The quantized field in spherical coordinates}

The choice of a particular vacuum state and the associated Fock representation of the quantized scalar field theory in de Sitter space is equivalent to the choice of a decomposition of the space $\mathcal{S}$ of solutions of the Klein-Gordon equation with smooth Cauchy conditions as a direct sum $\mathcal{S}=\mathcal{S}^+ \oplus \mathcal{S}^-$ of subspaces of positive and negative energy. In this section we construct the space $\mathcal{S}$ and describe the decomposition associated with the Bunch--Davies vacuum. Canonical quantization based on such a decomposition is then briefly discussed. In the last subsection, the one-particle subspace of the Fock representation is interpreted as an irreducible representation of de Sitter group, and explicit expressions for the generators of the group and discrete symmetries are written. We consider de Sitter radii and particle masses compatible with principal and with complementary series representations.

\subsection{Normal modes}

\label{normal-modes}

We start our analysis describing in some detail the normal modes of the Klein-Gordon equation for the $2d$ de Sitter space. This is required for the process of cannonical quantization we will present later.

The simplest way of looking at the $2d$ de Sitter space $dS^2$ is to consider it a submanifold embedded in a 3d Minkowski space $M^3$. Choosing a metric $\eta_{ab} = \textrm{diag}(1,-1,-1)$ for $M^3$, one has
\[
dS^2 = \{\vect X \in M^3 \mid \vect X^2 = X^a X^b \eta_{ab} = - \alpha^2 \} \, ,
\]
where $\alpha > 0$ is the de Sitter radius. The space so obtained is a hyperboloid, with topology $S^1 \times \mathbb{R}$. One may think of it as a spatial circle evolving in time. The symmetry group is the de Sitter group $O(2,\; 1)$, i.e., the isometries are Lorentz transformations in the ambient space. Changing to the so-called spherical coordinates,
\begin{align*}
X^0 & = \alpha \sinh(t/\alpha) \, , \\
X^1 & = \alpha \cosh(t/\alpha) \cos \theta \, , \\
X^2 & = \alpha \cosh(t/\alpha) \sin \theta \, ,
\end{align*}
the geometry of $dS^2$ is described by the induced Lorentzian metric tensor, with components:
\[
g_{00} = 1 \, , \quad g_{01} = 0 \, , \quad g_{11} = - \alpha^2 \cosh^2(t/ \alpha) \, .
\]
The volume density is $\sqrt{-g} = \alpha \cosh(t/\alpha)$, and the D'Alembertian is
\[
\square = \partial_{tt} + \frac{1}{\alpha} \tanh(t/\alpha) \partial_t - \frac{1}{\alpha^2 \cosh^2(t/ \alpha)} \partial_{\theta \theta} \, .
\]
The Klein-Gordon equation reads
\begin{equation}
\left( \square - \frac{m^2 + \xi R}{\hbar^2} \right) \phi = 0 \, .
\label{eq:klein-gordon}
\end{equation}
The scalar curvature is related to the de Sitter radius by $R =  2 / \alpha^2$. We put $\mu^2 = m^2 + \xi R$.

After separation of variables, the Klein-Gordon equation becomes
\begin{gather*}
\psi''= - k^2 \psi \quad \Rightarrow \quad \psi_k(\theta) = \frac{1}{\sqrt{2 \pi}}\textrm{e}^{i k \theta} \, , \quad k \in \mathbb{Z} \, ,  \\
T'' + \frac{1}{\alpha} \tanh(t/ \alpha) T' + \left( \frac{\mu^2}{\hbar^2} + \frac{k^2}{\alpha^2 \cosh^2(t/\alpha)} \right) T = 0  \, .
\end{gather*}
In order to solve the time-dependence of these ``angular momentum modes'' described by the index $k$, put $x = i \sinh(t/\alpha)$, and get:
\begin{equation}
(1 - x^2) \frac{d^2 T}{dx^2} - 2x \frac{dT}{dx} + \left[ - \frac{\alpha^2 \mu^2}{\hbar^2} - \frac{k^2}{1-x^2} \right] T = 0 \, .
\label{eq:time-eq-legendre}
\end{equation}
This is an associated Legendre equation. The solutions are associated Legendre functions $P_\nu^k(x)$ and $Q_\nu^k(x)$, with $\nu(\nu + 1)=- \alpha^2 \mu^2 / \hbar^2$. The coefficient $\nu$ is given by
\begin{equation}
\nu = \frac{- 1 \pm \sqrt{1- 4 \alpha^2 \mu^2 / \hbar^2}}{2} \, .
\label{eq:baskara-nu}
\end{equation}
If $\mu^2$ is positive, then $\nu$ is either a real number in the interval $[-1,0]$ or a complex number with real part equals to $-1/2$ and some nonzero imaginary part. If $\mu^2 = 0$, then $\nu=0,\; 1$. If $\mu^2$ is negative, then $\nu$ may assume arbitrary real values.

Throughout this work we will restrict to the case $\mu^2 > 0$. The squared mass is always positive, so this restriction corresponds in fact to not allowing a large negative coupling with the scalar curvature. In this case, a nice pair of linearly independent solutions of \eqref{eq:time-eq-legendre} is given by
\[
T^k_\nu\big(i \sinh(t /\alpha)\big) \, , \quad T^k_\nu\big(-i \sinh(t/\alpha)\big) \, ,
\]
where $T^k_\nu(z):=\textrm{e}^{\mp ik\pi / 2} P^k_\nu(z)$ for $\pm\mbox{Im}(z)>0$ is the Legendre function in `Ferrer's notation' \cite{snow}. The function $T^k_\nu(z)$ is analytic in the whole complex plane, except for two branching cuts on the real axis, one from $-\infty$ to $-1$ and another from $+1$ to $+\infty$. It doesn't matter which root $\nu$ of \eqref{eq:baskara-nu} is taken: both give the same function (that follows from the symmetry $T^k_\nu = T^k_{-\nu-1}$). The functions $T^k_\nu(z)$ have the property that $[T^k_\nu(i x)]^* = T^k_\nu(- i x)$, i.e., the linearly independent solutions are complex conjugate (see Appendix \ref{complex-conjugate-modes}).

Thus, there is a set of normal modes of the Klein-Gordon equation in de Sitter space \eqref{eq:klein-gordon} of the form:
\begin{align}
u_k(t,\theta) & = \sqrt{\frac{\gamma_k}{2}} \, T^k_\nu \big(i \sinh(t/\alpha)\big) \frac{\textrm{e}^{i k \theta}}{\sqrt{2\pi}}  \, , \nonumber \\
v_k(t,\theta) & = \sqrt{\frac{\gamma_k}{2}} \, T^k_\nu\big(-i \sinh(t/\alpha)\big) \frac{\textrm{e}^{i k \theta}}{\sqrt{2\pi}} \, ,
\label{eq:normal-modes}
\end{align}
with $k \in \mathbb{Z}$, and where $\gamma_k  := \Gamma(-\nu-k) \Gamma(\nu-k+1)$ are conveniently chosen normalization coefficients, which will be discussed latter. Here, $\Gamma$ is Euler's gamma function.

\subsection{Space of solutions and positive-energy modes}

\label{space-S}

The normal modes derived in the last section can be used for the construction of the space $\mathcal{S}$ of complex solutions of the Klein-Gordon equation in de Sitter space \eqref{eq:klein-gordon} with smooth Cauchy conditions. It is the vector space formed by wavefunctions of the form:
\begin{equation}
\phi(t,\theta)  \; = \; \sum_{k=-\infty}^{\infty} \big[ c_k u_k(t,\theta) + d_k v_k(t,\theta) \big] \, , 
\label{eq:solution-space}
\end{equation}
with coefficients $c_k,\; d_k$ of rapid decay in $|k|$, such as $\sum|c_k| |k|^l < \infty, \forall l>0$, and similarly for $d_k$. In order to see that, let us first prove that the series converges absolutely and uniformly to a $C^2$-solution of the Klein-Gordon equation with smooth initial conditions. Consider the sum containing terms with $c_k$ first. From \eqref{eq:normal-modes}, one has at each fixed $t$ a Fourier series:
\begin{align}
&C(t,\theta)  := \sum_{k=-\infty}^{\infty} p_k \frac{\textrm{e}^{i k \theta}}{\sqrt{2\pi}}\, , 
\label{eq:sum-C}\\
&p_k :=c_k \sqrt{\frac{\gamma_k}{2}} \,T^k_\nu \big(i \sinh(t/\alpha)\big) \, . \nonumber
\end{align}
The asymptotic behavior of the coefficients $p_k$ can be obtained from the large $k$ asymptotic representation of the Legendre functions $T_\nu^k$ given in Eq.~VI.95b of \cite{snow}. One finds that
\begin{align*}
\left| \sqrt{\gamma_k} \, T^k_\nu\big(i \sinh(t/\alpha)\big) \right| & \; =  \; \left| \frac{\Gamma(-\nu-k)}{\Gamma( \nu-k+1)} \right|^{1/2} \left|\Gamma( \nu-k+1) T^k_\nu(i \sinh(t/\alpha)\big) \right| \\
 \; & \; \simeq |k|^{-1/2} \, , \qquad \textrm{for large } k \textrm{ and }\forall t  \; .
\end{align*}
Since the $c_k$ are of rapid decay, the series \eqref{eq:sum-C} converges absolutely and uniformly in spacetime. Similar arguments can be used for the terms with coefficients $d_k$. Thus, the sum in Eq.~\eqref{eq:solution-space} is uniformly convergent. It also follows that $\phi(t,\theta)$ is continuous, since all terms in the uniformly convergent series are continuous. Consider now the derivatives $\partial_\theta^m \partial_t^n \phi$. A finite number $n$ of spatial derivatives changes the coefficients $p_k$ by a factor $(ik)^n$, and the time derivatives have the following form for large $k$:
\begin{align}
\frac{d}{dt} T^k_\nu\big(i \sinh(t/\alpha)\big)  &  \; \simeq \;  \frac{-ik}{\alpha \cosh(t/\alpha)} T^k_\nu\big(i \sinh(t/\alpha)\big) \, , \nonumber \\
\frac{d^2}{dt^2} T^k_\nu\big(i \sinh(t/\alpha)\big) &  \; \simeq  \; \left[- \frac{ik}{\alpha^2} \frac{\sinh(t/\alpha)}{\cosh^2(t/\alpha)} -  \frac{k^2}{\alpha^2 \cosh^2(t/\alpha)} \right] T^k_\nu\big(i \sinh(t/\alpha)\big) \, ,
\label{eq:asymptotic-time-derivatives}
\end{align}
thus changing the Fourier coefficients only polynomially in $k$, and in a uniform manner in $t$. But then, rapid decay of the $c_k$'s ensure uniform convergence of all second derivatives of \eqref{eq:sum-C}. Again, all arguments can be repeated for the sum involving the $d_k$'s. It follows that $\phi(t,\theta)$ is $C^2$ and, by construction, a solution of the Klein-Gordon equation (derivatives can be applied inside the sum, and each term is a solution of the equation). Moreover, at each fixed $t=t_0$, the restriction $\phi(t_0,\theta)$ is a smooth function on the circle, since any number of spatial derivatives can be applied to \eqref{eq:sum-C}. From Eq.~\eqref{eq:asymptotic-time-derivatives}, the same is true for $\dot{\phi}(t_0,\theta)$. We will therefore consider only smooth Cauchy data  $\phi(0,\theta),\; \dot{\phi}(0,\theta)$. Finally, any smooth function on the circle has a Fourier series with rapidly decaying coefficients, so $\mathcal{S}$ contains all solutions with smooth initial coefficients.

The vector space of solutions $\mathcal{S}$ is equipped with an invariant Hermitian sesquilinear form:
\begin{equation}
\scalar{f}{g} = i a(t) \int_{S_t} \textrm{d}\theta (f^* \partial_t g - \partial_t f^* g) \, ,
\label{eq:inv-scalar-prod}
\end{equation}
where $S_t$ is any spatial slice of constant time $t$, and $a(t): = \alpha \cosh(t/\alpha)$ is the corresponding scale factor --the radius of the circle $S_t$. Notice that $\scalar{f^*}{g^*}=-\scalar{g}{f}$. Our choice of the normalization coefficients $\gamma_k$ in Eq.~\eqref{eq:normal-modes} ensures that the normal modes are orthonormal,
\begin{equation}
\scalar{u_k}{u_l} = \delta_{kl} \, , \qquad \scalar{v_k}{v_l} = - \delta_{kl} \, , \qquad \scalar{u_k}{v_l} = 0 \, . 
\label{eq:uv-scalar-products}
\end{equation}
Let us prove that. The fact that $\scalar{u_k}{v_l} = 0$ is a simple consequence of the definition of the invariant form together with the identity $[T^k_\nu(i x)]^* = T^k_\nu(- i x)$. For the modes $u_k$ (put now $y= \sinh(t/\alpha)$),
\[
\scalar{u_k}{u_l} = - \delta_{kl} \frac{1}{2} |\gamma_k| \cosh^2(t/\alpha) \left[ T^k_\nu(- i y) \, T^{k \, \prime}_\nu(i y) + \textrm{c.c.} \right] \, .
\]
Invoke now the identity (from \cite{snow})
\begin{equation}
(1-z^2) \left[ T^k_\nu(z) \, \frac{d}{dz} T^k_\nu(-z) - \frac{d}{dz} T^k_\nu(z) \, T^k_\nu(-z) \right] = \frac{2}{\gamma_k} 
\label{eq:Tk-jacobian}
\end{equation}
in order to get
\[
\scalar{u_k}{u_l} = \delta_{kl} \frac{|\gamma_k|}{\gamma_k} \; .
\]
Let us show that $\gamma_k$ is positive for any $k$. Consider first the case $k=0$. Then
\[
\frac{1}{\Gamma(-\nu) \Gamma(\nu +1)} = \frac{\sin[(\nu+1)\pi]}{\pi} \, ,
\]
where $\nu$ is either a real number in the interval $(-1,\;0)$, or a complex number of the form $-1/2 +i \lambda$, $\lambda \in \mathbb{R}$. In the first case, $\nu+1$ is in the interval $(0,\;1)$, so $\sin[(\nu+1)\pi]$ is positive. In the second case, $\sin[(\nu+1)\pi] = \cosh(\pi \lambda)$, positive too. For general $k$, first note that, for $k$ positive,
\begin{equation*}
\frac{1}{\gamma_k} \; =\; \frac{ \prod_{l=1}^k (-\nu - l) (\nu - l + 1)}{\Gamma(-\nu) \Gamma(\nu +1)} 
\; = \; \frac{1}{\gamma_0} \prod_{l=1}^k(\alpha^2 \mu^2/\hbar^2 + l^2 -l)\, .
\end{equation*}
It is clear that the product is positive ($l^2 \geq l$ when $l$ is integer). A similar trick does the work for negative $k$. That completes the proof that $\scalar{u_k}{u_l} = \delta_{kl}$. For the case of the normal modes $v_k$, one can use the identity
\begin{equation}
\sqrt{\gamma_k} \, T_\nu^k(z) = \sqrt{\gamma_{-k}} \, T_\nu^{-k}(z)
\label{eq:gammas-rel}
\end{equation}
(which is proved using the inversion formula for gamma functions and Eq.~\eqref{eq:invert-k}) to see that $v_k = u^*_{-k}$. Then the result obtained for the modes $u_k$ implies that $\scalar{v_k}{v_l} = - \delta_{kl}$.

In order to proceed to the canonical quantization of the scalar field $\phi(t,\theta)$, one needs to choose a special decomposition of the space of solutions $\mathcal{S}$ as a direct sum $\mathcal{S}=\mathcal{S}^+ \oplus \mathcal{S}^-$, where $\mathcal{S}^+$ ($\mathcal{S}^-$) is interpreted as the space of positive (negative) energy solutions. The decomposition must be such that: (i) positive energy solutions have positive norm and (ii) the complex conjugate of a positive energy solution is a negative energy solution. From previous results, these conditions are satisfied if one picks a basis $\{u_k(t,\theta); k \in \mathbb{Z}\}$ for $\mathcal{S}^+$ and a basis $\{v_k(t,\theta); k \in \mathbb{Z}\}$ for $\mathcal{S}^-$. This is the energy-splitting decomposition that will be used in this work. There are alternative valid decompositions which, after canonical quantization, are associated with distinct choices of the vacuum state of the quantized theory. Our choice will lead to the so-called Bunch--Davies vacuum, in which we are interested due to its invariance under de Sitter group actions and its Hadamard property.

\subsection{One-particle subspace and canonical quantization} 

\label{canonical-quantization}

The Newton-Wigner (NW) position operator will be defined in the so-called ``one-particle subspace'' $\mathcal{H}$. This Hilbert space is defined as the completion of $\mathcal{S}^+$ in the scalar product defined by the sesquilinear form~\eqref{eq:inv-scalar-prod} (which is positive when restricted to $\mathcal{S}^+$). The vectors $\phi \in \mathcal{H}$ are superpositions of positive-energy solutions, and can be represented explicitly as
\begin{equation}
\phi(t,\theta) = \sum_k \phi_k u_k(t,\theta) \, , \qquad \sum_k |\phi_k|^2 = 1 \, , \qquad \phi_k \in \mathbb{C} \, .
\label{eq:mode-expansion}
\end{equation}
The scalar product in $\mathcal{H}$ is simply $\scalar{\phi}{\psi} = \sum \phi_k^* \psi_k$. We are going to think of the one-particle subspace as describing the quantum dynamics of a single relativistic particle in de Sitter space, following the usual physical interpretation: $\phi(t,\theta)$ will be the spacetime representation of the wavefunction associated with the particle. Some problems with this interpretation might be expected---it has been repeatedly remarked that the concept of particle for quantum fields in curved spacetimes is not well-defined. Nevertheless, it is just as clear that there are situations where a particle-like behavior is evident. As remarked in \cite{fewster}, particle physics experiments are actually performed in a curved spacetime, and we do see particle tracks in experiments. To understand how to deal with a quantum field theory in a curved spacetime under circumstances where a particle-like behavior is possible is one of the purposes of this paper.

After fixing the positive-energy modes of the classical field, canonical quantization in de Sitter space follows pretty much the same steps as in Minkowski space \cite{birrel-davies}, as we now briefly describe. One considers the bosonic Fock space $\mathcal{F}$ built in the usual way from the one-particle state $\mathcal{H}$ as $\mathcal{F}:= \mathbb{C} \oplus_{n=1}^{\infty} \left(\mathcal{H}^{\otimes n}\right)_{s}$, where the index ``$s$'' indicates the symmetrization of the tensor products. Following the usual prescription, the quantized neutral massive scalar field, acting in $\mathcal{F}$, is expressed in the form
\begin{equation}
\hat{\phi}(t,\theta) = \sum_{k=-\infty}^{\infty} (a_k u_k + a_k^* u_k^*) \, ,
\label{eq:mode-expansion-real}
\end{equation}
where the $u_k$ are the chosen orthonormal positive energy modes, and the $a_k,\; a_k^*$ are annihilation and creation operators satisfying the commutation relations
\[
[a_k,\, a_l^*] = \delta_{kl} \, , \qquad [a_k,\, a_l] = [a_k^*,\, a_l^*] = 0 \, .
\]
The vacuum $\ket{\Omega}$ is defined as the vector state annihilated by all annihilation operators, $a_k \ket{\Omega} = 0$, $\forall k$, and many-particle states are created by repeated application of creation operators to the vacuum.

It is a well known fact that the quantization of free fields in curved space times is non-unique, and different choices of positive energy decompositions may lead to unitarily inequivalent representations of the algebra of cannonical commutation relations. These different choices reflect  different possible choices for the vacuum state.

Our particular choice of the previously defined positive energy modes $u_k$ in the expansion (\ref{eq:mode-expansion-real}) corresponds to the choice of the so called ``Bunch--Davies vacuum''. In order to establish this claim, we present an explicit calculation of the two-point function in Appendix \ref{two-point}, and compare it to the two-point function obtained in the original work of Bunch and Davies \cite{bunch-davies} (where flat coordinates were used), showing that both results agree. The choice of the Bunch--Davies vacuum is particularly relevant because of its previously mentioned relation to the Hadamard condition \cite{allen,BFH}.

\subsection{Group action on the space of positive-energy solutions}

\label{group-action}

The space $\mathcal{S}^+$ of positive-energy solutions was described in a given system of spherical coordinates $(t,\theta)$, but there is a whole family of systems $(t', \theta')$ related by isometries in the de Sitter group $O(2,\;1)$. We want to prove that the definition of $\mathcal{S}^+$ is coordinate-independent, i.e., that the subspace of positive-energy modes is invariant under the action of the group (what is equivalent to the invariance of the vacuum state in the quantized theory), as well as to find out how the group acts on these modes. That will lead to an interpretation of its completion $\mathcal{H}$ as an irreducible representation of de Sitter group.

Any element of $O(2,\; 1)$ is the product of an element of the restricted de Sitter group $O(2,\;1)_+^\uparrow$ of Lorentz transformations of determinant $1$ which do not reverse the direction of time, and possibly parity $\mathbf{P}$ and time reversal $\mathbf{T}$. There are three linearly independent generators in the algebra of $O(2,\;1)$, which may be taken as the infinitesimal boosts along the rectangular axes, $N_{10}$ and $N_{20}$, and the generator of rotations, $N_{12}$. The question is how these transformations act on the modes defined in Eq.~\eqref{eq:normal-modes}. The case of rotations is quite simple. A transformation $U_{12}(\phi)= \exp(\phi N_{12})$ which rotates the space by an angle $\phi$ changes angles in spherical coordinates according to $\theta \mapsto \theta - \phi$, while the coordinate $t$ remains unaffected. The generator of rotations is $N_{12} = - \partial/ \partial \theta$. Its action on the basis vectors is just $N_{12} \, u_k = - i k \, u_k$, i.e., the basis $\{ u_k \}$ is that of the eigenvectors of the Hermitian operator $i N_{12}$.

Now consider the case of $N_{10}$. Since Lorentz transformations are naturally described in the flat coordinates of the ambient Minkowski space $M$, let us describe the modes $u_k$ in the same coordinates:
\[
u_k = \sqrt{\frac{ \gamma_k}{4 \pi}} \, T^k_\nu(- i X^0 / \alpha) \, \frac{(X^1 + i X^2)^k }{[\alpha^2 + (X^0)^2]^{k/2} } \, .
\]
An infinitesimal Lorentz transformation along the axis $X^1$ is given by
\begin{align*}
(X^0)' & = X^0 - \lambda X^1 \, , \\
(X^1)' & = X^1 - \lambda X^0 \, , \\
(X^2)' & = X^2 \, ,
\end{align*}
where $\lambda$ is the infinitesimal parameter of the transformation (the transformation is Lorentz to first order in $\lambda$). Thus, the variation of $u_k$ is
\[
N_{10} \, u_k \; = \; X^1\frac{\partial u_k}{\partial X^0} + X^0 \frac{\partial u_k}{\partial X^1}  \, .
\]
A similar equation holds for boosts along the axis $X^2$. Evaluating the derivatives and using a few relations between Legendre functions from \cite{snow}, one finds that the action of the generators of de Sitter group is
\begin{align}
N_{12} u_k & = - i k u_k \, , \nonumber \\
N_{10} u_k & = \frac{i}{2} |(\nu+k)(\nu - k + 1)|^{1/2} \, u_{k-1} \nonumber \\
	& \qquad + \frac{i}{2} |(\nu-k)(\nu + k + 1)|^{1/2} \, u_{k+1} \, , \label{eq:dS-algebra-H} \\
N_{20} u_k & = \frac{1}{2} |(\nu+k)(\nu - k + 1)|^{1/2} \, u_{k-1} \nonumber \\
	& \qquad - \frac{1}{2} |(\nu-k)(\nu + k + 1)|^{1/2} \, u_{k+1} \, . \nonumber
\end{align}
These equations show that $\mathcal{H}$ is closed under the action of the infinitesimal generators. Hence, $\mathcal{H}$ is a representation space for $O(2,\;1)_+^\uparrow$, the action of a Lorentz transformation $L$ on a wavefunction $\phi(x) \in \mathcal{H}$ being given by $\phi(x) \mapsto \phi(L^{-1}x)$. The Casimir operator which characterizes the irreducible representations is $C= N_{12}^2 - N_{10}^2 - N_{20}^2$, and is easily verified to be $C = - \nu(\nu+1)=\alpha^2\mu^2/\hbar^2$ for the above expressions.

With our restriction to $\mu^2 > 0$, the index $\nu$ may be: (a) a real number in the interval $(-1,\;0)$; or (b) a complex number of the form $\nu=-1/2 + i \lambda$, with $\lambda \in \mathbb{R}$. In the case (a), one has $0< C < 1/4$, what corresponds to a representation of de Sitter group in the so-called complementary series (the continuous representations $C_q^0$ in the exceptional interval $0 < q < 1/4$ in Bargmann's work \cite{bargmann}). In the case (b), one has $C \geq 1/4$, what corresponds to principal series representations (continuous representations $C_q^0$ with $q \geq 1/4$).

Now let us introduce the discrete symmetries of parity $\mathbf{P}$ and time-reversal $\mathbf{T}$. We represent parity as the reversal of the axis $X^2$ in the ambient Minkowski space. Then parity just reverses the sign of the angular coordinate of a wavefunction in $\mathcal{H}$, $\mathbf{P} \phi(t,\theta) = \phi(t,-\theta)$. In particular, for the basis vectors $u_k$, one may use the identity  \eqref{eq:gammas-rel} in order to get
\begin{equation}
\mathbf{P} u_k = u_{-k} \, .
\label{eq:parity-H}
\end{equation}
The action of $\mathbf{T}$ has a peculiarity connected with the restriction to the space of positive-energy states. The geometrical realization of the transformation is the reversal of the time coordinate in the ambient Minkowski space. But this cannot be represented as $\phi(t,\theta) \mapsto \phi(-t,\theta)$, since the result is a negative-energy state. In order that the transformation is closed in $\mathcal{H}$, we take the anti-unitary representation $\mathbf{T} \phi(t,\theta) = \phi^*(-t,\theta)$. But then the action of the operator on modes $u_k$ is the same as that of parity, with the difference that the action is anti-linear,
\begin{equation}
\mathbf{T} u_k = u_{-k} \qquad \textrm{(anti-linear)} \, .
\label{eq:time-reversal-H}
\end{equation}

\section{Newton-Wigner localization}

\subsection{Definition of the localization system}

\label{sec:def-localization}

The notion of localization of relativistic particles in Minkowski space provided by the Newton-Wigner (NW) position operator was introduced in \cite{newton-wigner}. In that paper, a list of properties is postulated, which are assumed to hold for any reasonable relativistic position operator, and it is proved that there is a unique operator satisfying them. A more direct way to understand this position operator is described in \cite{haag}. Let us review the basic argument. Consider a massive scalar field in Minkowski space. The one-particle subspace of the theory consists of vectors $\phi(p) \in L^2(\mathbb{R}, dp/\omega(p))$, with $\omega(p) = \sqrt{p^2 + m^2}$, i.e., the scalar product is
\[
\scalar{\phi}{\psi} = \int \frac{ \textrm{d}p}{\omega(p)} \phi^*(p) \psi(p) \, .
\]
Now absorb a factor $\sqrt{\omega(p)}$ in each wavefunction: i.e., consider the unitary transformation $M_\omega: L^2(\mathbb{R}, dp/\omega(p))\to L^2(\mathbb{R}, dp)$, whose action is $\phi(p) \mapsto \phi_{NW}(p)=\phi(p)/\sqrt{\omega(p)}$. Then, introduce a unitary operator of time-evolution $U_t :L^2(\mathbb{R}, dp) \to L^2(\mathbb{R}, dp)$, represented by the transformation $\phi_{NW}(p) \mapsto (U_t \, \phi_{NW})(p) = \exp(- i \omega(p) t / \hbar) \phi_{NW}(p)$. Finally, Fourier transform the result in order to get a spatial representation,
\[
\phi_{NW}(t,x) = \frac{1}{\sqrt{2 \pi}} \int \textrm{d}p \, \textrm{e}^{ipx/\hbar } \textrm{e}^{- i \omega(p) t/\hbar }\, \phi_{NW}(p)  \, .
\]
That gives the Newton-Wigner wavefunction. The probability density that the particle is detected at the point $x$ in time $t$ is $P(t,x)=|\phi_{NW}(t,x)|^2$. The position operator itself, at time $t$, is the multiplication operator in the spatial representation at the same time,
\[
(q_t \phi)_{NW}(t,x) = x \phi_{NW}(t,x) \, .
\]

Some difficulties show up if one tries to repeat the same steps in the case of de Sitter space. First, there is no canonical definition of a momentum space representation. We overcome this problem by looking at the mode expansion as a convenient (for our purposes) de Sitter analogue of the Fourier transform. It is clear that a mode expansion is a coordinate dependent concept, therefore the resulting position operator will depend on the choice of coordinates. But, as well known,  the Newton-Wigner operator is not a covariant object even in Minkowski space: there is a distinct operator associated with each reference frame. The problem found in Minkowski space is just carried over into de Sitter space, and we do not attempt to solve it here.

The second point is the absence of a time-translation isometry in $dS^2$, what makes the time-evolution of individual modes much more complicated than in Minkowski space. Two aspects are relevant here: there is no definite frequency $\omega_k$ associated with each mode, so that time-evolution in momentum space is not just multiplication by varying phases $\exp(-i \omega t / \hbar)$ as before; and the oscillation of the field goes on together with a damping of the field amplitude, forced by the expansion of the universe (for increasing $|t|$). We will see that these effects can be isolated: the damping factor will be analogous to the factor $\sqrt{\omega}$ absorbed in the definition of the Newton-Wigner wavefunction in the Minlowski case, while the oscillating phases will be responsible for the time-evolution of the position operator.

Let us now proceed to the definition of the de Sitter version of NW-localization. Later we will interpret the results drawing an analogy with the discussion above. We assume that a localization system in de Sitter space is:

\begin{description}

\item[I] A family of unitary transformations $W_t:\mathcal{H} \to L^2(S^1)$, $\phi \mapsto \phi_{NW}(t,\theta)$, where $L^2(S^1)$ is the Hilbert space of square-integrable functions on the circle $S^1$;

\item[II] If $U_{12}(\alpha) \in \textrm{SO}(1,\;2)$ is a rotation by an angle $\alpha$, then $U_{12}(\alpha) \phi \mapsto \phi_{NW}(t,\theta-\alpha)$;

\item[III] $\mathbf{P} \phi \mapsto \phi_{NW}(t,-\theta)$, and $\mathbf{T} \phi \mapsto \phi_{NW}^*(- t,\theta)$;

\item[IV] In the large mass limit, one has $\phi_{NW}(t,\theta) \,\propto\, \phi(t,\theta)$.

\end{description}
Condition IV will be clarified below. As a regularity condition, we also assume that $W_t^{-1}:L^2(S^1) \to \mathcal{H}$ depends continuously on the mass $m$. Let us discuss the intuitive content of the Postulates above.

The Newton-Wigner wavefunction $\phi_{NW}(t,\theta)$ is interpreted, for each time $t$, as describing quantum amplitudes for finding the particle at position $\theta$. In other words, the probability of finding the particle in a Lebesgue measurable set $I$ is $P(I) = \int_I |\phi_{NW}(t,\theta)|^2 d\theta$. Postulate I corresponds to the basic requirement that such a probability distribution exists for each time $t$.

The second postulate is that the Newton-Wigner representation is well-behaved under rotations. A rotation in de Sitter group, when seen from the Newton-Wigner spatial representation, must rotate the probability amplitudes on the circle by the same angle. This condition can be reformulated as $W_t U_{12}(\varphi) W_t^* = R(\varphi)$, where $R$ is the operator of rotation for square-integrable functions on the circle.

Postulate III is the requirement that the discrete symmetries of parity and time-reversal act as geometrical transformations on the Newton-Wigner representations. The complex conjugation in the time-reversal condition is necessary because the image of an anti-unitary operator under a unitary equivalence must be anti-unitary too. A quantum symmetry is in general defined up to a phase, according to the celebrated Wigner's theorem; we are assuming here that the phases are equal to $1$, avoiding complications with the possibility of a projective representation of the extended de Sitter group.

Finally, the Postulate IV is necessary in order to fix some remaining ambiguities in $W_t$, as we shall see. It is motivated by the following fact. In the large mass limit, the scalar product \eqref{eq:inv-scalar-prod} of one-particle states $\phi(t,\theta),\;\psi(t,\theta)$ reduces to
\begin{equation}
\scalar{\phi}{\psi} \; \simeq \; 2 \frac{\mu a(t)}{\hbar} \int_{S_t} d\theta \, \phi^*(t,\theta) \psi(t,\theta) \, ,
\label{eq:scalar-product-large-m}
\end{equation}
i.e., it becomes the scalar product of $L^2$ functions on the circle. In this case, it is natural to interpret $|\phi(t,\theta)|^2$ directly as a  probability distribution (up to the factor outside the integral). The postulate IV ensures that the Newton-Wigner distribution agrees with such an interpretation.

The consequences of the postulates can now be evaluated. Let us start with postulate II. For each $t$, a suitable basis for $L^2(S^1)$ is that composed of eigenvectors of the Hermitian generator of rotations. That is the same as describing the NW-wavefunction in its Fourier expanded form,
\[
\phi_{NW}(t,\theta) = \sum_k q_k(t) \frac{\textrm{e}^{i k \theta}}{\sqrt{2\pi}} \, , \qquad \sum_k |q_k(t)|^2= 1 \, ,
\]
with $q_k \in \mathbb{C}$. Consider the vector $u_k \in \mathcal{H}$. The action of a rotation $U_{12}(\alpha)$ on it is to multiply the state by a phase, $U_{12}(\alpha) u_k = \exp(- i k \alpha) u_k$. Since $U_{12}$ is linear, the same must be true for its image in $L^2(S1)$:
\[
W_t\left( U_{12}(\alpha) u_k \right) = \textrm{e}^{- i k \alpha} W_t(u_k) \, ,
\]
what implies
\begin{equation}
R(\alpha) W_t(u_k) = \textrm{e}^{-i k \alpha} W_t(u_k) \, .
\label{eq:rot-covariance}
\end{equation}
But then it must be
\begin{equation}
W_t(u_k) = \textrm{e}^{- i \varphi_k(t)} \, \frac{\textrm{e}^{i k \theta}}{\sqrt{2\pi}} \, ,
\label{eq:condition-rotation}
\end{equation}
where $\varphi_k(t)$ is some arbitrary phase. For, suppose the space $V$ of solutions $W_t(u_k)$ of Eq.~\eqref{eq:rot-covariance} has more than one dimension. Note that the action of the Hermitian generator $J$ of rotations in $V$ is multiplication by $k$. Then there would be at least two orthogonal vectors with the same eigenvalue $k$, what is impossible, since the eigenspaces of $J$ are non-degenerate. Therefore, $V$ is one-dimensional, the space of eigenvectors of $J$ with eigenvalue $k$. Because the transformation $W_t$ is unitary, and $u_k$ has norm $1$, there is just a phase freedom, what corresponds to Eq.~\eqref{eq:condition-rotation}.

The action of parity in $\mathcal{H}$ is given by Eq.~\eqref{eq:parity-H}. The first part of postulate III, when applied to the general form of the solution of postulate II described in Eq.~\eqref{eq:condition-rotation}, leads to
\begin{equation}
W_t(u_k) = W_t(\mathbf{P} u_{-k}) = \textrm{e}^{- i \varphi_{-k}(t)} \, \frac{\textrm{e}^{i k \theta}}{\sqrt{2\pi}} \, .
\label{eq:condition-parity}
\end{equation}
The action of time-reversal in $\mathcal{H}$ is given by Eq.~\eqref{eq:time-reversal-H}. The second part of postulate III leads to
\begin{equation}
W_t(u_k) = W_t(\mathbf{T} u_{-k}) = \textrm{e}^{i \varphi_{-k}(-t)} \, \frac{\textrm{e}^{i k \theta}}{\sqrt{2\pi}} \, ,
\label{eq:condition-time-reversal}
\end{equation}
Put $s_k(t) := \textrm{e}^{- i\varphi_k(t)}$. Comparing Eqs.~\eqref{eq:condition-rotation}, \eqref{eq:condition-parity} and \eqref{eq:condition-time-reversal}, one finds that
\begin{equation}
s_k(t) = s_{-k}(t) \, , \qquad s_k(t) = s_k^*(-t) \, .
\label{eq:sign-ambiguity}
\end{equation}
The form of the transformation $W_t$ is restricted, but not uniquely fixed by the axioms I--III. The varying phases must satisfy Eq.~\eqref{eq:sign-ambiguity}, but there remains a lot of freedom after these conditions are imposed. In the next section we describe the additional restrictions which follow from postulate IV at $t=0$, where a unique solution is obtained. Then we study the case of generic $t$, and suggest a natural solution based on an analogy to the case of Minkowski space.

\subsection{The case of $t=0$}

At $t=0$, the identities in Eq.~\eqref{eq:sign-ambiguity} simplify to $s_k(0)=s_{-k}(0)=\pm 1$. Thus, the transformation $W_0$ determined by postulates I--III is given by:
\begin{equation}
\phi(0,\theta) = \sum_k \phi_k \sqrt{\frac{\gamma_k}{2}} T_\nu^k(0) \frac{\textrm{e}^{i k \theta}}{\sqrt{2\pi}} \mapsto \phi_{NW}(0,\theta) = \sum_k \phi_k \, s_k(0) \frac{\textrm{e}^{i k \theta}}{\sqrt{2\pi}} \, .
\label{eq:NW-time-0}
\end{equation}
As we see, there is a sign ambiguity in each term of the above series, due to the presence of the factor $s_k(0)$. 

This ambiguity was first pointed by Philips and Wigner in \cite{philips-wigner}. Below, we will discuss how these authors address this problem, but let us first show that our postulate IV fixes these ambiguities in a more natural way, leading to a unique solution for $W_0$.

From Eq.~\eqref{eq:baskara-nu} and the definition $\mu^2 = m^2+\xi R$, it follows that the large mass limit $m \to \infty$ corresponds to the limit $\lambda \to \infty$ in the index of the Legendre functions $\nu = -1/2 + \lambda i$. In order that the postulate IV is satisfied, it is necessary that $\phi_{NW}(0,\theta) = f(m) \phi(0,\theta)$ in Eq.~\eqref{eq:NW-time-0}, with some mass-dependent normalization factor $f(m)$. But from the asymptotic expression for the Legendre function in the large $\nu$ limit (see Eq.~VI.(93a) in \cite{snow}), and using the Stirling approximation for Gamma functions, one gets
\begin{equation}
\sqrt{\frac{\gamma_k}{2}} T_\nu^k(0) \; \simeq \;  (-1)^k (2 \lambda)^{-1/2} \, .
\label{eq:large-mass-gamma}
\end{equation}
The factor $(2 \lambda)^{-1/2}$ is just a mass-dependent normalization coefficient, as can be seen from Eq.~\eqref{eq:scalar-product-large-m} with $a(0)=\alpha$, since $\lambda \to \alpha \mu / \hbar$ for large masses. Therefore,  postulate IV implies that, up to an irrelevant overall sign, one has
\begin{equation}
s_k(0) \; = \;(-1)^k \, .
\label{eq:sk}
\end{equation}

The fact that the same choice is made for all masses $m$ is a consequence of the asumption that $W_0^{-1}$ is continuous with respect to the mass $m$. Since $s_k(0)=\pm 1$, it cannot change but discontinuously. Summing up: the transformation $W_0$ is completely determined by the postulates I-IV, being given by Eq.~\eqref{eq:NW-time-0} with $s_k(0)=(-1)^k$.

\subsection{Heuristical discussion on the ambiguities of signs}

Let us discuss some  heuristical aspects of the position probability distribution we have found and the origin of the sign ambiguities occuring in the coefficients $s_k(0)$ before the postulate IV is used. As we shall now explain, the existence of the sign ambiguities is a consequence of the fact that the localization postulates I-III alone do not fix the localization in the one-particle space $\mathcal{H}$ relative to the  localization in the Newton-Wigner representation space $L^2(S^1)$.

Consider a specific choice of the signs $s_k=\pm 1$. If these coefficients are changed according to $s_k \mapsto s_k^\prime = (-1)^k$, then the Newton-Wigner representation is rotated by an angle of $\pi$. Hence, part of the freedom in the choice of the $s_k$ is due to the possibility of applying a rotation of $\pi$. In fact, analizing the action of the map $W_0$ on a sufficiently large class of carefully chosen states, it is even possible to fix all sign ambiguities just by avoiding such antipodal reflections.

In order to see this, consider a simple example. Take a superposition of $k=-1,0,1$ states with $a_0=s_0$, $a_1=a_{-1}=s_1 /2$:
\begin{eqnarray}
\phi(0,\theta) & = &\sqrt{\frac{\gamma_0}{4\pi}} \, T_\nu^0(0) s_0 + \sqrt{\frac{\gamma_1}{4 \pi}} \, T_\nu^1(0) s_1 \cos \theta \;,
\label{eq:Localizacao-1}
\\
\phi_{NW}(0,\theta) & = &\frac{1 + \cos \theta}{\sqrt{2\pi}} \, .
\label{eq:Localizacao-2}
\end{eqnarray}
The Newton-Wigner wavefunction (\ref{eq:Localizacao-2}) has a maximum at $\theta=0$ and decreases monotonically with increasing $|\theta|$, assuming the value zero at the antipodal point $\theta=\pi$. That is, it describes a particle more likely to be found in the region $|\theta|<\pi/2$ than in the antipodal related region. Let us now consider behavior of the corresponding one-particle state (\ref{eq:Localizacao-1}). One has
\begin{equation}
T_\nu^k(0)= \frac{(-1)^k 2^k \sqrt{\pi}}{\Gamma\left(\frac{\nu-k}{2}+1\right) \Gamma\left(\frac{-\nu-k+1}{2}\right)} 
\label{eq:T-nu-zero}
\end{equation}
and it can be proved that the product of $\Gamma$'s in the denominator is  positive. Thus, the  coefficients $T_\nu^k(0)$ have signs alternating in $k$, because of the factor $(-1)^k$. There are two distinct possibilities for the action of $W_0$: either $s_0=s_1$ or $s_0 \neq s_1$. If $s_0=s_1$, then the two terms in the r.h.s.\  of (\ref{eq:Localizacao-1}) have different signs at $\theta=0$, and the same sign at $\theta=\pi$, and the wavefunction has higher amplitudes in the region $\pi/2 < |\theta|<\pi$, with its maximum at $\theta=\pi$. On the other hand, if $s_0 \neq s_1$, then $\phi(0,\theta)$ has its maximum at $\theta=0$, and higher amplitudes in the region $|\theta|<\pi/2$. The two choices are related by a rotation of $\pi$. Therefore, in order that the wavefunction $\phi(0,\theta)$ is concentrated at the same region as $\phi_{NW}(0,\theta)$, and not at the antipodal related region, one must choose $s_0 \neq s_1$.

The same argument can be adapted to states constructed by superpositions of states with $|k|=p,\; p+1$, allowing one to fix $s_{p+1}$ in terms of $s_p$. Notice that this argument works only for very special states: in a generic state $\phi$,  distinct choices of $s_k(0)$ are related by more complicated transformations than simple rotations by $\pi$.

In any case, as we saw above, different choices of signs reflect on the localization of states in the one-particle space $\mathcal{H}$ relative to the Newton-Wigner representation spece $L^2(S^1)$. According to the usual interpretation of the localization operators and of the wavefunctions, it is natural to choose the signs in a way that the same interpretation of localization is found in both spaces: if the Newton-Wigner wavefunction is concentrated about some $\theta_0$, the corresponding one-particle state should be concentrated at the same region, and not at the antipodal point.

The choice $s_k(0)=(-1)^k$ can be obtained, alternatively, from a condition of ``maximal localization'' of position eigenstates. Consider a sequence of localized functions in the Newton-Wigner representation,
\begin{equation}
\delta_{NW}^K(\theta) = \frac{1}{\sqrt{2\pi}} \sum_{|k|<K} \frac{\textrm{e}^{i k \theta}}{\sqrt{2\pi}} \, , \quad K \in \mathbb{N} \, ,
\label{eq:delta-sequence}
\end{equation}
which converges, for $K \to \infty$ (in a distributional sense), to the Dirac delta function $\delta_{NW}(\theta)$. Allowing for the sign freedom in the coefficients $s_k(0)$, these functions correspond to one-particle states of the form:
\begin{equation}
\delta^{K}(0,\theta) = \frac{1}{\sqrt{2\pi}} \sum_{|k|<K} s_k(0) \sqrt{\frac{\gamma_k}{2}} T_\nu^k(0) \frac{\textrm{e}^{i k \theta}}{\sqrt{2\pi}} \, ,
\label{eq:delta-sequence-st}
\end{equation}
so that at $\theta=0$ one has:
\[
\delta^{K}(0,0) = \frac{1}{2\pi} \sum_{|k|<K} s_k(0) \sqrt{\frac{\gamma_k}{2}} T_\nu^k(0) \, .
\]
As we saw from Eq.~\eqref{eq:T-nu-zero}, we have $\textrm{sign}\,\big(T_\nu^k(0)\big) = (-1)^k$. Hence, the choice $s_k(0)=(-1)^k$ maximizes the value of $|\delta^{K}(0,0)|$, for all $K$. In other words, that is the choice which makes the localized state $\delta_{NW}(\theta)$ in the Newton-Wigner representation as concentrated as possible about $\theta=0$ in the spacetime representation at $t=0$. The notion of localizability contained in the transformation $W_0$ is associated with maximally localized wavefunctions being well-behaved under de Sitter group symmetries.

\subsection{Comparison with Philips-Wigner states}

An earlier discussion of localizability in de Sitter space was presented by Philips and Wigner in \cite{philips-wigner}, and we would like to compare our results to theirs. A brief review of \cite{philips-wigner} is presented in Appendix \ref{sec:pw}, to which we refer for more details. The main result obtained in that paper was a description of (improper) states $\phi^{(\theta_0)}$ localized at position $\theta_0$ in $t=0$. Such states were described in terms of their Fourier coefficients in an explicit principal series representation of de Sitter group on a space $L^2(S^1)$ of square-integrable functions on the circle. We want to compare these with the localized states $\eta^{(\theta_0)}=\delta_{NW}(\theta-\theta_0)$ of our NW-representation in $t=0$, whose Fourier coefficients in the NW-representation are given by $\exp(- i k \theta_0)/\sqrt{2\pi}$. From Eqs.~\eqref{eq:NW-time-0} and \eqref{eq:sk}, these localized states correspond to (improper) one-particle states in $\mathcal{H}$ of the form:
\begin{equation}
\eta^{(\theta_0)}(t,\theta) = \sum_k \frac{(-1)^k}{\sqrt{2 \pi}} \textrm{e}^{- i k \theta_0} \sqrt{\frac{\gamma_k}{2}} T_\nu^k\big(i \sinh (t/\alpha)\big) \frac{\textrm{e}^{i k \theta}}{\sqrt{2\pi}} \, ,
\label{eq:localized-states-spacetime}
\end{equation}
that is, they have components
\begin{equation}
\eta^{(\theta_0)}_k = \frac{(-1)^k}{\sqrt{2 \pi}} \textrm{e}^{- i k \theta_0}
\label{eq:eta-localized-states}
\end{equation}
in $\mathcal{H}$. We shall restrict in this section to representations of the principal series, on which the work \cite{philips-wigner} is based.

Let us start by discussing the relation between the representation of Sitter algebra in the space $\mathcal{H}$ of one-particle states described in Section \ref{group-action} and the more traditional Bargmann's representations used in \cite{philips-wigner}. The principal series Bargmann representation on $\mathcal{H}' := L^2(S^1)$ is briefly reviewed in Appendix \ref{sec:pw}. Let $\{\ket{k}\}$ be the basis of $\mathcal{H}'$ composed of normalized eigenstates of the generator of rotations, $\ket{k} = \exp(i k \theta)/\sqrt{2 \pi}$. The action of de Sitter algebra in this basis is given by:
\begin{align}
N_{12} \ket{k} & = - i k \ket{k} \, , \nonumber \\
N_{10} \ket{k} & = \frac{\nu + k}{2} \ket{k-1} + \frac{\nu - k}{2} \ket{k+1} \, , \label{eq:dS-algebra-pw}\\
N_{20} \ket{k} & = i \frac{\nu + k}{2} \ket{k-1} - i \frac{\nu - k}{2} \ket{k+1} \, , \nonumber
\end{align}
with $\nu = -1/2 + \lambda i$. These expressions are direct translations of Eqs.~\eqref{eq:PW-rotations} and \eqref{eq:PW-boosts}, discussed in more detail in \cite{philips-wigner}. On the other hand, the representation of de Sitter algebra on $\mathcal{H}$ is described explicitly in Eq.~\eqref{eq:dS-algebra-H}. The principal series representations are those with $\nu = -1/2 + \lambda i$, $\lambda \in \mathbb{R}$, in which case Eq.~\eqref{eq:dS-algebra-H} reduces to the simpler form:
\begin{align}
N_{12} u_k & = - i k u_k \, , \nonumber \\
N_{10} u_k & = \frac{i}{2} |\nu+k| u_{k-1} + \frac{i}{2} |\nu-k| u_{k+1} \, , \label{eq:dS-algebra-H-ps} \\
N_{20} u_k & = - \frac{1}{2} |\nu+k| u_{k-1} + \frac{1}{2} |\nu-k| u_{k+1} \, . \nonumber
\end{align}
Comparing the expressions in Eqs. \eqref{eq:dS-algebra-pw} and \eqref{eq:dS-algebra-H-ps}, it can be verified that a unitary equivalence $U_B^\dagger: \mathcal{H}' \to \mathcal{H}$ is given by $\ket{k} \mapsto \chi_k u_k$, where $\chi_k$ is a complex number defined by the recurrence relations:
\[
\chi_0 =1 \, , \qquad \chi_{k+1} = -i \frac{\nu+k+1}{|\nu+k+1|} \chi_k \, .
\]
The last relation is equivalent to
\[
\chi_{k-1} = -i \frac{\nu-k+1}{|\nu-k+1|} \chi_k \, .
\]
Hence, the coefficients can be written as ($k>0$):
\begin{equation}
\chi_k = \chi_{-k} = (-i)^k \frac{\left(\frac{1}{2} + i \lambda \right) \left(\frac{3}{2} + i \lambda \right) \dots \left( k - \frac{1}{2} + i \lambda \right)}{\left|\left(\frac{1}{2} + i \lambda \right) \left(\frac{3}{2} + i \lambda \right) \dots \left( k - \frac{1}{2} + i \lambda \right)\right|} \, .
\label{eq:change-basis}
\end{equation}
Now let $\mathcal{H}_{NW}^{(0)} := L^2(S^1)$ denote the space of Newton-Wigner wavefunctions at time $t=0$, and build the composition $U_B^{(0)} := U_B \circ W_0^\dagger$. This transformation maps a NW-wavefunction to the corresponding state in Bargmann's representation. Choosing the basis $\big\{\ket{k} = \exp(i k \theta)/\sqrt{2 \pi}, \; k\in\mathbb{Z}\big\}$ for $\mathcal{H}_{NW}^{(0)}$, one has $W^\dagger_0 \ket{k}= (-1)^k u_k$. Therefore, a NW-wavefunction $\phi_{NW}(0,\theta) = \sum \phi_k \ket{k}$ corresponds to a vector $\phi_B(\theta)= \sum (-1)^k \chi_k^* \phi_k \ket{k}$ in Bargmann's representation.

A Philips-Wigner state $\phi^{(\pi/2)}$ localized at $\theta = \pi/2$ has Fourier coefficients $\sqrt{2 \pi} l_k$ in $\mathcal{H}'$ given by the explicit formula displayed in Eq.~\eqref{eq:pw-loc-coefficients}. Such state can be mapped to $\mathcal{H}$ with the help of the unitary transformation $U_B^\dagger$. One finds that $U_B^\dagger \, \phi^{(\pi/2)}$ has coefficients $\phi_k^{(\pi/2)} = \sqrt{2 \pi} l_k \chi_k$ in $\mathcal{H}$. From Eq.~\eqref{eq:change-basis} and Eq.~\eqref{eq:pw-loc-coefficients}, it follows that $l_k \chi_k = i^k$. Now compare with our localized states. From Eq.~\eqref{eq:eta-localized-states}, a NW-state $\eta^{(\pi/2)}$ localized at $\theta_0=\pi/2$ at $t=0$ has coefficients $\eta_k^{(\pi/2)} = i^k/\sqrt{2 \pi}$. Therefore, up to an irrelevant normalization factor, the localized states at $\theta = \pi /2$ are the same. Besides, both classes of states behave in the same way under rotations. So our definition of localized states allows one, for $t=0$, and with the restriction to the principal series, to recover the results of \cite{philips-wigner}.

Therefore, working in a quite different setting, we have arrived at the same localized states as obtained in \cite{philips-wigner} in the context of group theory. We would like to discuss now the main differences and similarities between these approaches, and emphasize some technical simplifications and a conceptual clarification we believe our work brings to the discussion of localizability in de Sitter spacetime.

Compare with what happens in Minkowski space. The work of Newton and Wigner \cite{newton-wigner} was written in terms of distributions describing improper states localized at specific points. The results were latter reformulated by Wightman \cite{wightman} in terms of projectors $E(S)$ in a Hilbert space associated with observables describing the property of the particle being in a region $S$ of space. So the idea of localization at a point was replaced by localization in a finite region. The main advantage in doing this is that technical complications associated with the theory of distributions are avoided. In short, the approach of Wightman allowed the results of \cite{newton-wigner} to be derived in a rigourous manner in the simplest context of a Hilbert space of square-integrable functions. In de Sitter space, the work of Philips and Wigner follows the original idea of looking for localized states, while we have studied the analogue of Wightman's localizability postulates, describing the probability $P(S)$ of detection of the particle in a finite, measurable region $S$ in terms of the norm of its state $\phi$ projected to a suitable subspace in the Hilbert space $\mathcal{H}$, i.e., $P(S)=\int_S d\theta |\phi_{NW}(\theta)|^2=\|\chi_S \, \phi\|^2$, where $\chi_S$ is the projection operator which acts as the characteristic function of $S$ in the Newton-Wigner representation.

In Minkowski space, the approaches of \cite{newton-wigner} and \cite{wightman} are essentially equivalent: the conditions used by Wightman were direct translations of the original conditions on distributions. In de Sitter spacetime, however, there are important differences between our approach and that of \cite{philips-wigner}. Firstly, distinct sets of axioms are used: one of the axioms of \cite{philips-wigner} (Axiom c in Appendix \ref{sec:pw}) is replaced in our approach by postulate IV. These conditions play similar roles in each approach, being required for the elimination of sign ambiguities encoded in the factors $s_k(0)=\pm 1$. However, the physical motivation of Axiom (c)---minimal disturbance of a localized state at $\theta_0$ under the action of boosts which leave the point $\theta_0$ invariant---is unclear. According to our previous discussion, the ambiguities can be fixed in a more natural way by a condition of `optimal localization'---that the probabilities should be as concentrated as possible about the wavefunction---and can be easily implemented requiring a reasonable large mass limit.

Moreover, there is a residual ambiguity in \cite{philips-wigner}. This is removed by requiring that the localized states have positive energy in the Minkowski space limit, what is done using a complicated process of contraction (roughly speaking, by taking the limit $\alpha\to\infty$) of Lie algebra representations of the de Sitter group.  This step is not necessary in our approach. Such a simplification is due to the fact that negative energy states are not allowed here, the starting point being a space of positive-energy solutions of the Klein-Gordon equation. Importantly, we employ the Hadamard condition to select states with positive energy, making it unnecessary to check the sign of the energy of the localized states through a Minkowski space limit.

A second remark concerns the connection between representations of de Sitter group and wave equations in de Sitter space. As widely known, the irreducible representations of de Sitter group were classified by Bargmann in \cite{bargmann}. Yet, when one considers applications to quantum field theory, it is natural to ask for an interpretation of the representations in spaces of solutions of wave equations in de Sitter space. More than that, one wants to restrict to positive-energy solutions. Here we have used the Hadamard condition in order to select a suitable space of positive-energy solutions of the Klein-Gordon equation, and displayed such an interpretation for the principal and complementary series representations.

In doing so, we identified the one-particle subspace of the quantized massive scalar field theory with a particular irreducible representation of the de Sitter group. In an intuitive sense, that identification provides a spacetime representation for vectors in the more abstract (from a physicist's perspective) Bargmann's representations: it allows one to see these vectors as wavefunctions in de Sitter space. In particular, it becomes possible to determine how the localized states are spread in spacetime, i.e., Eq.~\eqref{eq:localized-states-spacetime} (remember that relativistic localized states are not strictly localized, but ``as localized as possible'' states). This question could not be dealt with without a prescription for the choice of the positive-energy states, and was not investigated in \cite{philips-wigner}.

In Minkowski space, if one wants to see how a localized state defined in momentum space looks like in spacetime, one just goes to configuration space, using the well-known transformation $\phi(\vect{p}) \mapsto \phi(\vect{x},t)$, the relativistic Fourier transform. The result is a Hankel function, with an exponential decay for large spatial distances \cite{newton-wigner}. Such a familiar  transformation has no natural analogue in curved spacetimes. A Bargmann's representation can be seen as a sort of `momentum representation', but the transformation to configuration space---that of wavefunctions in de Sitter space---is not unique: it depends on the choice of a vacuum, or equivalently, of the positive-energy states. It is necessary to combine purely group theoretical results with the modern specification of positive-energy states given by the Hadamard condition in order to find the spacetime representation of states of interest.

\subsection{Time-evolution of the Newton-Wigner wavefunction}

The postulates I--IV determine uniquely the form of the Newton-Wigner wavefunction at time $t=0$. They also impose restrictions on the time-evolution of the wavefunction, but do not fix it uniquely. In this section we discuss a solution of these conditions, suggested by an analogy with the definition of the position operator in Minkowski space discussed in the beginning of Section \ref{sec:def-localization}. It is natural that in generalizing structures defined in Minkowski space to the context of curved spacetimes some non-uniqueness might be met with. Nevertheless, one would certainly like to restrict it as much as possible, and in de Sitter space there is the advantage of dealing with a maximally symmetric spacetime. We discuss later in this section the possibility of using the group symmetry operations in de Sitter space in order to fix the Newton-Wigner dynamics, but we will answer this question in the negative, at least for a simple implementation of these symmetries.

So, let us describe a solution of the time-evolution $W_t$ of the Newton-Wigner wavefunction. Keep in mind the discussion in Section \ref{sec:def-localization}. From Eq.~\eqref{eq:mode-expansion} and the explicit form of the normal modes given in Eq.~\eqref{eq:normal-modes}, a generic vector in $\mathcal{H}$ can be written as
\begin{equation}
\phi(t,\theta) = \sum_k \phi_k \sqrt{\frac{\gamma_k}{2}} \, T^k_\nu\big(i \sinh(t/\alpha)\big) \frac{\textrm{e}^{i k \theta}}{\sqrt{2\pi}} \, ,
\label{eq:general-state}
\end{equation}
with $\sum_k |\phi_k|^2 = 1$. The scalar product is given by Eq.~\eqref{eq:inv-scalar-prod}, which reduces to $\scalar{\phi}{\psi} = \sum \phi_k^* \psi_k$ in this representation. Now, introduce
\[
N_k(t) :=  \frac{1}{\gamma_k \, \left| T^k_\nu\big(i \sinh(t/\alpha)\big) \right|^2}  \, ,
\]
and
\begin{equation}
\varphi_k(t) := - \arg \Bigl( T^k_\nu\big(i \sinh(t/\alpha)\big) \Bigr) 
\label{eq:time-evolution-phase}
\end{equation}
and define a time-dependent unitary transformation $W_t$ given for each $t$ by
\begin{equation}
\phi(t,\theta) \mapsto \phi_{NW}(t,\theta) = \sum_k \phi_k \textrm{e}^{- i \varphi_k(t)} \, \frac{\textrm{e}^{i k \theta}}{\sqrt{2\pi}} \, .
\label{eq:nw-rep}
\end{equation}
Above, a factor $[2 N_k(t)]^{-1/2}$ is absorbed in each coefficient $\phi_k$, and a time-evolution $\exp[- i \varphi_k(t)]$ is associated with each mode. The analogy with the definition of the position operator in Minkowski space should be clear. The absorption of the factor $[2 N_k(t)]^{-1/2}$ is a consequence of postulates I--III; what is added is the choice of the phases $\varphi_k(t)$ prescribed by Eq.~\eqref{eq:time-evolution-phase}. Note that for $t=0$, it follows from Eq.~\eqref{eq:T-nu-zero} and Eq.~\eqref{eq:time-evolution-phase} that $\exp(-i \varphi_k(0))=(-1)^k$. Substituting that in Eq.~\eqref{eq:nw-rep}, we get the operator $W_0$ obtained in the previous section. Hence, we have the right transformation at $t=0$. Moreover, it is easy to check that the postulates I--IV are satisfied. It must be verified that $\textrm{e}^{-i \varphi_{k}(t)} = \textrm{e}^{-i \varphi_{-k}(t)}$ (parity), $\textrm{e}^{-i \varphi_k(t)} = \textrm{e}^{i \varphi_k(-t)}$ (time-reversal), and that the large mass limit is correct. That the parity condition is satisfied is a consequence of Eq.~\eqref{eq:gammas-rel}, which shows that $T_\nu^{-k}(z)$ and $T_\nu^k(z)$ are proportional, with a positive proportionality factor. The time-reversal condition is a consequence of the identity $\Big[T^k_\nu\big(i \sinh(t/\alpha)\big)\Big]^* = T^k_\nu\big(-i\sinh(t/\alpha)\big)$. And the large mass limit can be established using Eq.~VI.(90) of \cite{snow}:
\[
P_\nu^\mu(z) \simeq \frac{\nu^{\mu-1/2}}{\sqrt{2 \pi}(z^2-1)^{1/4}} \left[ - \textrm{e}^{\pm i (\mu-1/2)\pi} \textrm{e}^{i(\nu+1/2)\omega} + \textrm{e}^{-i(\nu+1/2) \omega} \right] \, , \quad \textrm{ for } \nu_2 \to \pm \infty \, ,
\]
where $\nu=\nu_1+i\nu_2$ and $z = \cos \omega$, together with the Stirling approximation for complex numbers with large imaginary part \cite{remmert}, which gives
\[
\left|\Gamma\left(\frac{1}{2}-k-\lambda i \right)\right| \simeq \sqrt{2 \pi} \lambda^{-k} \textrm{e}^{-\pi \lambda/2} \, ,
\]
so that the analogue of Eq.~\eqref{eq:large-mass-gamma} is
\[
\sqrt{\frac{\gamma_k}{2}} T_\nu^k\big(i \sinh (t/\alpha)\big) \simeq (-1)^k (2 \lambda \cosh \big(t/\alpha)\big)^{-1/2} \textrm{e}^{-i \lambda t/\alpha}
\]
for large $m$. The factor $\big(2 \lambda \cosh (t/\alpha)\big)^{1/2}$ is a mass-dependent normalization coefficient, from Eq.~\eqref{eq:scalar-product-large-m}.

There is a simple physical interpretation for the prescribed choice of the phases $\varphi_k(t)$. The Newton-Wigner wavefunction at time $t$ is described in Eq.~\eqref{eq:nw-rep} as a square-integrable function on a circle of radius $1$. Its squared value gives the probability of finding the particle in an infinitesimal interval of angles. But the actual spatial radius of the corresponding time slice is the scale factor $a(t) = \alpha \cosh (t/\alpha)$, so if one wants to get the probability density in the spatial slice itself, a factor of $\sqrt{a(t)}$ must be included, leading to
\[
\tilde{\phi}_{NW}(t,\theta) = \frac{1}{\sqrt{\alpha \cosh (t/\alpha)}} \sum_k \phi_k \textrm{e}^{- i \varphi_k(t)} \, \frac{\textrm{e}^{i k \theta}}{\sqrt{2\pi}} \, .
\]
In this case, the transformation which defines $\tilde{\phi}_{NW}(t,\theta)$ involves the absorption of a factor $[2 \omega_k^{dS}(t)]^{-1/2}$, with
\begin{equation}
\omega_k^{dS}(t) :=  \frac{1}{\gamma_k \left| T^k_\nu\big(i \sinh(t/\alpha)\big) \right|^2 \alpha \cosh(t/\alpha)}  \, .
\label{eq:dispersion-relation}
\end{equation}
There is an interesting relation between the derivative of the phases $\varphi_k(t)$ and the factors $\omega_k^{dS}(t)$. Pick Eq.~\eqref{eq:Tk-jacobian} and consider it on the imaginary axis, with $z= iy$. Divide it by $\left| T^k_\nu(- i y) \right|^2 = T^k_\nu(i y) \, T^k_\nu(- i y)$:
\begin{align}
\frac{- 2}{\gamma_k (1+y^2) \left| T^k_\nu(- i y) \right|^2} & = \left[ \frac{T^{k \, \prime}_\nu(-i y)}{T^k_\nu(- i y)} + \frac{T^{k \, \prime}_\nu(i y)}{T^k_\nu(i y)} \right]  \nonumber \\
& = 2 \textrm{Re}\left[ \frac{T^{k \, \prime}_\nu(i y)}{T^k_\nu(i y)} \right] \, .
\label{eq:phases-derivative-aux}
\end{align}
The last identity follows from the fact that the derivative of a Legendre function can be written as a (real) linear combination of Legendre functions, which are complex conjugated by the inversion $iy \to - iy$. Now, Eq.~\eqref{eq:time-evolution-phase} implies 
\[
\varphi_k^\prime(t) = - \frac{1}{\alpha} \cosh(t/\alpha) \, \textrm{Re} \left[ \frac{T^{k \, \prime}_\nu\big(i \sinh(t/\alpha)\big)}{T^k_\nu\big(i \sinh(t/\alpha)\big)} \right] \, ,
\]
what in turn, using Eq.~\eqref{eq:dispersion-relation} and Eq.~\eqref{eq:phases-derivative-aux}, leads to
\[
\varphi_k^\prime(t) = \omega_k^{dS}(t) \, .
\]
Therefore, the dynamics described by Eq.~\eqref{eq:nw-rep} corresponds to that generated by normal modes $k$ with a time-dependent energy $\omega_k^{dS}(t)$, with the time $t=0$ representation fixed by postulates I--IV. In other words, we are looking at Eq.~\eqref{eq:dispersion-relation} as a time-dependent dispersion relation giving the energy of a mode $k$ as a function of time.

The solution we have found for the localizability postulates was written in a form valid both for the principal and complementary series, i.e. Eq.~\eqref{eq:nw-rep}. But it must be noticed that for representations in the complementary series the normal modes $T_\nu^k\big(i\sinh(t/\alpha)\big)$ are purely real and not oscillatory, implying that the phases $\exp(-i\varphi_k(t))$ are constant. Therefore, the dynamics of the Newton-Wigner function in this case is trivial; there is no time-evolution. Notice that complementary series representations are associated with particle masses and de Sitter radii satisfying $4 \alpha^2 \mu^2 / \hbar^2 < 1$. This inequality essentially states that the Compton wavelength of the scalar particle is bigger than the radius of de Sitter spacetime. Under these circumstances the NW-wavefunction describing the position of the particle is not able to move, and the notion of localization is trivial. In the principal series representations, on the other hand, the Newton-Wigner function has a nontrivial dynamics, and the time-evolution of the distribution of probabilities might be used in order to study how wavepackets move.

It is a well-known fact that the Newton-Wigner operator displays acausal features in the propagation of wavepackets in Minkowski space, i.e. superluminal propagation is possible \cite{ruijsenaars}, and the same phenomenon should be expected in the case of de Sitter spacetime. The central result of \cite{ruijsenaars} was a determination of upper bounds for the probability of detection of superluminal propagation. These depend on the experimental techniques available, but using some generous estimates of experimental parameters it was found in \cite{ruijsenaars} that the probability of detecting superluminal propagation in a single experiment is smaller than $10^{-10^8}$. It was also argued that a more carefull exam should reduce this bound considerably. It is natural to expect that curvature-dependent corrections to this result should be present in de Sitter spacetime, but given the order of magnitude of the effect, there shall be a considerable window in the space of parameters ($m$, $\alpha$, a typical time-scale $T$, etc) where acausal effects are negligible.

For particles in the principal series, the position operator should be useful for the study of the dynamics of relativistic particles in de Sitter spacetime in a semiclassical regime. One could prepare an initially localized wavepacket, and study how the probability density propagates and spreads. It is clear that the expansion of the universe will enforce an additional spreading in the dispersion of wavepackets compared to that in Minkowski space, so that a wavepacket resembling a localized particle will remain so only for a finite amount of time. But that is a general feature in the study of classical limits of quantum systems, true even for non-relativistic systems: typically a classical limit is approached only inside a finite interval of time, as discussed in the classical work \cite{hepp}. For how long the classical limit is reasonable, and how exactly the wavepackets spread is described by the time-evolution of the position probability distribution.

Finally, we should briefly mention that, following a suggestion made in \cite{philips-wigner}, but not further developed there, we analysed the possibility of deriving the dynamics of the NW-wavefunction from the action of the de Sitter group on the NW-wavefunctions at time $t=0$, thus trying to remediate with boots and rotations the absence of time-translation isometries. However, our attempts led to incompatibilities with our more natural postulates I--III and we came to the conclusion that the implementation of this seemingly sound idea is actually quite subtle, perhaps impossible.

\section{Perspectives}

We have showed that a notion of localization exists for massive neutral scalar fields in de Sitter space compatible with the prescription for the choice of positive energy modes encoded in the Hadamard condition. In de Sitter space, this condition is equivalent to the choice of the Bunch--Davies vacuum as the ``physical vacuum'' among the family of $\alpha$-vacua. Therefore, we have proved that localizability is compatible with this choice of vacuum. A natural question arises whether other choices of vacuum are compatible or not with localizability. If they are not, that would be another argument in favor of the Bunch--Davies vacuum. We expect to investigate this problem in a future work.

Another direction of research is related to the problem of understanding the classical limit of quantum field theories in curved spacetimes. Following the general procedure for studying classical limits introduced by Hepp in \cite{hepp}, we have proved in a previous work \cite{mcl} that the quantum theory of the free neutral massive scalar field in Minkowski space has two distinct kinds of classical limits: one of them describing a classical field theory, the other one a classical particle dynamics. The Newton-Wigner position operator is used in order to prove the existence of the latter. We expect that the same problem can be investigated in de Sitter space along similar lines, with the position probability distributions discussed herein playing the role of the Newton-Wigner operator.

We have considered particles with positive mass in the principal and complementary series of representations of de Sitter group. In the case of the complementary series, the position operator was found to be trivial, without dynamics. Hence, a nontrivial classical limit should exist only for representations in the principal series. Moreover, it has been recently discovered that it is also possible to formulate sensible free quantum fields in de Sitter space using representations with a negative mass, the so-called tachyonic fields of \cite{bros-2}. The question of the localizability of these fields was not treated here, and could be investigated in a future work.

\begin{acknowledgments}

This work was supported by FAPESP under Grant 2007/55450-0. NY thanks the Erwin Schr\"odinger Institute, Vienna, for support and hospitality during the program \textit{``Quantum Field Theory on Curved Spacetimes and Curved Target Spaces''}. He also thanks U.\ Moschella for discussions during the initial stages of his studies of quantum fields in de Sitter spacetime. Both authors specially wish to thank the referees for important remarks on a previous version of this manuscript that led to considerable improvements in our results.

\end{acknowledgments}

\appendix

\section{Proof of $[T^k_\nu(i y)]^* = T^k_\nu(- i y)$} 

\label{complex-conjugate-modes}

The functions $T^k_\nu(z)$ are defined for $|z-1|<2$ in terms of hypergeometric functions by
\begin{equation}
T^k_\nu(z) := \frac{ (1 - z^2)^{k/2} \Gamma(\nu + k +1)}{2^k \Gamma(k+1) \Gamma(\nu - k + 1)} \, f(z) \, ,
\label{eq:P-def-F}
\end{equation}
with
\[
f(z) := F\left( -\nu + k, \nu + k + 1, k + 1; \frac{1-z}{2} \right) \, .
\]

Let us see what happens to the function under complex conjugation. For $k \geq 0$, the hypergeometric function can be represented as a convergent power series in the radius $|z|<1$. The $(j+1)$-th term in the expansion of $f$ in powers of $(1-z)/2$ has a coefficient of the form
\[
\frac{\prod_{l=0}^j (-\nu + k + l) (\nu + k + 1 + l)}{\prod_{l=0}^j (k + 1 + l)} \, .
\]
The denominator is real, so ignore it. Recall that $\nu$ is a root of the quadratic equation $\nu(\nu+1) = - \alpha^2 \mu^2$, so whenever $\nu(\nu + 1)$ makes an appearance, it is a real number. It follows that every factor in the product is real. Thus, the power function has real coefficients, and $[f(z)]^* = f(z^*)$.

Now, the gamma functions. The factor $\Gamma(k+1)$ is real. The part that matters is
\[
\frac{ \Gamma(\nu + k +1)}{\Gamma(\nu - k + 1)} = (\nu+k) (\nu+k-1) \cdots (\nu - k +2) (\nu - k +1) \, .
\]
This can be rewritten as
\[
\prod_{l=0}^{k-1}  (\nu + k - l) (\nu + 1- k +l) = \prod_{l=0}^{k-1} \, [ \nu (\nu + 1) - (k-l)^2 + (k-l)] \, ,
\]
which is also real. Besides that, each factor in the product is a negative number: $\nu (\nu + 1) = - \alpha^2 \mu^2$, and $(k-l)^2 \geq (k-l)$, since $k-l$ is an integer. Therefore, the product is negative for odd $k$, and positive for even $k$. This result will be needed somewhere else.

Finally, take $z = iy$, $y \in \mathbb{R}$, and consider the factor $(1 - z^2)^{k/2}$. Here one must be careful. The functions $T_\mu^k$ are defined with square roots cut along distinct lines: the factor $\sqrt{1-z}$ is cut along $x>1$ on the real axis, while the factor $\sqrt{1+z}$ has a cut along $x<-1$. With these choices, $\sqrt{1-(ix)^2} = |1+iy|$. Then it follows that $\{[1 - (ix)^2]^{k/2}\}^* = [1 - (-ix)^2]^{k/2}$, so that $[T^k_\nu(i y)]^* = T^k_\nu((i y)^*) = T^k_\nu(- i y)$, at least in the radius $|z|<1$ and for $k \geq 0$.

In order to extend the result to the domain of $T^k_\nu$, introduce an auxiliary analytic function  $[T^k_\nu(-x,y)]^*$. This function coincides with $T^k_\nu(- z)$ along the imaginary axis inside the radius $|z|<1$. Moreover, both functions are defined on the same domain: $T^k_\nu(z)$ is single-valued on a domain invariant both under inversion $z \to -z$, and inversion of the real part $(x,y) \to (-x,y)$. Hence, $[T^k_\nu(-x,y)]^* = T^k_\nu(- z)$. Restricting to the imaginary axis, $[T^k_\nu(i y)]^* = T^k_\nu(- iy)$. The result is extended to negative $k$ using the relation
\begin{equation}
T_\nu^{-k}(z) = (-1)^k \frac{\Gamma(\nu - k + 1)}{\Gamma(\nu + k + 1)} T_\nu^{k}(z) \, .
\label{eq:invert-k}
\end{equation}
It was already proved that the factor with the $\Gamma$'s is real.

\section{Two-point function}

\label{two-point}

The two-point function, $G:= \bra{\Omega}\phi(t,\theta),\phi(t', 0)\ket{\Omega}$, is given by:
\begin{align*}
G & = \sum_k u_k(t,\theta) u_k^*(t',0) \\
				& = \sum_k \frac{\Gamma(-\nu-k) \Gamma(\nu - k +1)}{4 \pi} \textrm{e}^{ik \theta} T^k_\nu(i y) T^k_\nu(- i y') \\
				& = \frac{1}{4 |\sin \nu \pi|} \sum_k (-)^k \frac{\Gamma(\nu - k +1)}{\Gamma(\nu + k +1)} \textrm{e}^{ik \theta} T^k_\nu(i y) T^k_\nu(- i y') ,
\end{align*}
with $y = \sinh(t/\alpha)$. Call the sum in the last line $S$. Using \eqref{eq:invert-k}, it can be written as
\begin{align*}
S & = \sum_{k=-\infty}^{0} \left[ \textrm{e}^{ik \theta} T^{-k}_\nu(i y) T^k_\nu(- i y') \right]  - T_\nu(i y) T_\nu(- i y')  \\
		& \qquad + \sum_{k=0}^\infty \left[ \textrm{e}^{ik \theta} T^k_\nu(i y) T^{-k}_\nu(- i y') \right] \\
	& = 2 \sum_{k=0}^\infty \cos(k\theta) T^k_\nu(i y) T^{-k}_\nu(- i y') - T_\nu(i y) T_\nu(- i y') \\
	& = 2 \sum_{k=0}^\infty \epsilon_k \frac{\Gamma(\nu - k +1)}{\Gamma(\nu + k +1)} \cos(k(\pi - \theta)) T^k_\nu(i y) T^{k}_\nu(- i y') \, ,
\end{align*}
where  $\epsilon_k = 1 - \delta_{k0}/2$, i.e., $\epsilon_k$ is $1$ except for $k=0$, when it is $1/2$. There is a nice summation theorem for Legendre functions \cite{snow},
\[
P_\nu \left( z_1 z_2 + \sqrt{1-z_1^2} \sqrt{1-z_2^2} \cos \theta \right) \; = \; 
	2 \sum_{k=0}^\infty \epsilon_k \frac{\Gamma(\nu - k +1)}{\Gamma(\nu + k +1)} \cos(k\theta) T^k_\nu(z_1) T^{k}_\nu(z_2) \, ,
\]
which leads to
\[
S = P_\nu \left( y y' - \sqrt{1-(iy)^2} \sqrt{1-(-iy')^2} \cos \theta \right) = P_\nu(Z) \, .
\]
The argument in the function $P_\nu(Z)$ can be written in invariant form:
\begin{align*}
Z 	& = \sinh(t/\alpha) \sinh(t'/\alpha) - \cosh(t/\alpha) \cosh(t'/\alpha) \cos \theta \\
	& = \alpha^{-2} X \cdot X' \, ,
\end{align*}
where $X$ is the vector in the Minkowski space $M^3$ corresponding to the point $(t,\theta)$ in de Sitter space, while $X'$ corresponds to the point $(t', 0)$. Collecting the calculations,
\[
G = G(Z) = \frac{1}{4 |\sin(\nu \pi)|} P_\nu(Z) \, .
\]
The Legendre function is singular at $Z = -1$, where a cut begins which extends along the real axis to $-\infty$. This value has a simple geometric interpretation. Recall that the causality relations on de Sitter hyperboloid are inherited from the Minkowski ambient space: two points $x,x'$ are space (light,time) related if their corresponding vectors are space (light,time)-like. In particular, light-like related vectors satisfy $(X-X')^2=0 \Rightarrow X \cdot X' = - \alpha^2$, so that $Z = -1$ in this case. In other words, the two-point function is singular on the light cone. This property is characteristic of the Bunch--Davies vacuum: any other choice of modes would lead to an additional singularity at the antipodal points of the light-cone. Besides that, one can write the Legendre function in terms of a hypergeometric function to get
\[
G(Z) = \frac{1}{4 |\sin(\nu \pi)|} \, F\left(\nu +1,-\nu, 1;\frac{1-Z}{2} \right) \, .
\]
Compare with the original Bunch and Davies work \cite{bunch-davies}. In their notation, a coefficient $\mu$ is introduced:
\[
\mu = \sqrt{\frac{1}{4} - 2 \xi -m^2 \alpha^2} \; ,
\]
in terms of which our $\nu$ becomes $\nu = -1/2 + \mu$. It is easy to check this relation. Our definition \eqref{eq:baskara-nu} of $\nu$ can be rewritten using $R = - 2 / \alpha^2$ as
\[
\nu = - \frac{1}{2} \pm \sqrt{\frac{1}{4} - 2 \xi - m^2 \alpha^2} \, .
\]
Moreover, $\sin(\nu \pi) = \sin((-1/2 + \mu) \pi) = (-1) \cos(\mu \pi)$, where $\mu \in (-1/2, 1/2)$ or is purely imaginary, so that $\cos(\mu \pi)$ is positive either way. Thus,
\[
G(Z) = \frac{1}{4} \sec(\mu \pi) \, F\left(\nu +1,-\nu, 1;\frac{1-Z}{2} \right) \, .
\]
That is just the expression in Eq. (2.13) for the two-point function in \cite{bunch-davies}.

\section{Philips and Wigner localized states}

\label{sec:pw}

Our work has a close relation with that of Philips and Wigner \cite{philips-wigner}, and for the sake of the reader, we present now a brief review of their little known article. Their purpose was to investigate how the existence of localized states is related to the condition of positivity of the energy. But it was not known at that time how to define positive energy states in curved spaces, so in order to check the sign of the energy of a given state it was necessary to study the limit where the geometry of de Sitter approached that of Minkowski space, what was done invoking a contraction of the group representation. Although the problem remains unsolved in general, it is now known that, at least in the case of spacetimes with a compact Cauchy surface, the Hadamard condition is sufficient to fix the ambiguity in the choice of the positive energy solutions \cite{wald}.

Let us describe the unitary representation of the de Sitter group used in \cite{philips-wigner}. We restrict to the case of $O(2,\;1)$ which is relevant here. Let $\mathcal{H}'$ be the set of square-integrable functions $\psi(\theta)$ on the unitary circle $S^1$ on the Euclidean plane $\mathbb{R}^2$. Extend these functions to the whole plane: $\psi(\theta) \mapsto f(\rho) \psi(\theta)$, where $\rho$ is the radius $\rho = \sqrt{(X^1)^2 + (X^2)^2}$, and $f(\rho)$ is a fixed function, smooth and square-integrable on the plane. Rotations are realized as rotations on the circle, i.e.,
\begin{equation}
N_{12} = - \partial/ \partial \theta \, ,
\label{eq:PW-rotations}
\end{equation}
and infinitesimal boosts are represented by
\begin{gather}
N_{10} = - \sin \theta \frac{\partial}{\partial \theta} + \nu \cos \theta \, , \nonumber \\
N_{20} = \cos \theta \frac{\partial}{\partial \theta} + \nu \sin \theta \, ,
\label{eq:PW-boosts}
\end{gather}
where $\nu = -1/2 + i \lambda$. The generators can be integrated to give finite boosts and rotations, so that there are unitary operators $U(S)$ corresponding to each element $S$ of the restricted de Sitter group. The parity operator $\mathbf{P}$, understood as the representation of the geometric operation $p$ of reversing the axis $X^1$ in the ambient Minkowski space, must satisfy the group relations up to some projective factor,
\[
\mathbf{P} U(S) = \omega(S) U(pSp) \mathbf{P} \, ,
\]
where $S$ is any Lorentz transformation in the restricted de Sitter group. But it can be proved that $\omega(S) = 1$, and that 
\[
\mathbf{P} \psi(X^1,X^2) = \pm \psi(-X^1,X^2) \, ,
\]
where the choice of the sign must be the same for all $\psi$. This choice is physically irrelevant, so just pick the sign $+1$. For the time-reversal operator $\mathbf{T}$, the group relations lead to essentially two possibilities, corresponding to a unitary $\mathbf{T}_u$ or an anti-unitary $\mathbf{T}_a$, given by
\[
\mathbf{T}_u \psi(X^1,X^2) = \pm \psi(-X^1, -X^2) \, ,
\]
and 
\begin{equation}
\mathbf{T}_a \psi(\theta) = \int K(\theta - \theta') \psi^*(\theta') d\theta' \, , 
\label{eq:T-op}
\end{equation}
where the kernel $K$ is given in Fourier expanded form by
\begin{equation}
K(\theta - \theta') = \sum a_k \textrm{e}^{i k (\theta - \theta')} \, , \quad \frac{a_{k+1}}{a_k} = - \frac{\frac{1}{2} + k - i \lambda}{\frac{1}{2} + k + i \lambda} \, ,
\label{eq:T-op-coefficients}
\end{equation}
with $a_0 = 1 / (2 \pi)$. The coefficients automatically satisfy $a_k = a_{-k}$. In order that $\mathbf{T}_a$ is uniquely defined, it is assumed that $\mathbf{T}^2 = 1$ (it could be $-1$), and that $\mathbf{T} \mathbf{P} = \mathbf{P} \mathbf{T}$ (there could be a phase difference).

The definition of the localized states is based on a set of three postulates, which represent the de Sitter version of the postulates of Newton and Wigner adopted in the case of Minkowski space \cite{newton-wigner}. The postulates are:

\begin{description}
\item[a] A localized state is invariant under reflections that leave the point of localization invariant.
\item[b] A rotation applied to a localized state gives a new localized state --the point of localization is just rotated accordingly.
\item[c] A boost which keeps the point of localization invariant changes the state as little as possible.
\end{description}

The first result is that the postulates cannot be satisfied with a unitary time-reversal operator. Hence, the existence of localized states implies that $\mathbf{T}$ is anti-unitary --it must be the $\mathbf{T}_a$ defined in Eqs. \eqref{eq:T-op}, \eqref{eq:T-op-coefficients}. In this case, the postulates are satisfied by two distinct sets of localized states.

Consider a state $\psi_1(\theta)$ localized at $\theta = \pi /2$ at $t=0$. It must be invariant under parity and time-reversal. Writing a Fourier expansion
\[
\psi_1(\theta) = \sum l_k \textrm{e}^{i k \theta} \, ,
\]
invariance under parity implies
\[
\quad l_{-k} = (-1)^k l_k \, ,
\]
while invariance under time-reversal leads to
\[
2 \pi a_k l_{-k}^* = l_k \, .
\]
Combining these results, and using \eqref{eq:T-op-coefficients}, it follows that
\[
\frac{l_{k+1}}{l_k} = \zeta_{k+1/2} \frac{\frac{1}{2} + k - i \lambda}{\left[ (\frac{1}{2} + k)^2 + \lambda^2 \right]^{1/2}} \, ,
\]
where the $\zeta$'s are real numbers satisfying
\[
\zeta_{k+1/2} \zeta_{-k-1/2} = 1 \, .
\]
Then condition (b), together with (c), which reduces here to minimal deformation under boosts along $X^1$, fixes $\zeta = 1$ or $\zeta = -1$. The first possibility is ruled out by looking what happens in the contraction of the de Sitter group representation to a representation of the inhomogeneous Lorentz group. The choice $\zeta = 1$ corresponds to a state of negative-energy in Minkowski space in this limit. So it must be $\zeta = -1$. The Fourier coefficients of $\psi_1$ are then completely determined, being given by ($k>0$),
\begin{gather}
l_k = (-1)^k \frac{\left(\frac{1}{2} - i \lambda \right) \left(\frac{3}{2} - i \lambda \right) \dots \left( k - \frac{1}{2} - i \lambda \right)}{\left|\left(\frac{1}{2} - i \lambda \right) \left(\frac{3}{2} - i \lambda \right) \dots \left( k - \frac{1}{2} - i \lambda \right) \right|} , \nonumber \\
l_0 = 1 	\label{eq:pw-loc-coefficients} \\
l_{-k} = (-1)^k \frac{\left( - \frac{1}{2} + i \lambda \right) \left(- \frac{3}{2} + i \lambda \right) \dots \left(- k + \frac{1}{2} + i \lambda \right)}{\left|\left( - \frac{1}{2} + i \lambda \right) \left(- \frac{3}{2} + i \lambda \right) \dots \left(- k + \frac{1}{2} + i \lambda \right)\right|}  \, . \nonumber
\end{gather}
States localized at other angles are obtained with the application of rotations.

It is curious that in this approach the condition (b) that localized states are well-behaved under rotations is not as important as its counterpart in Minkowski space. It is necessary to supplement it here with the auxiliary condition (c), which has a more obscure interpretation --it is not an invariance condition, nor a mapping of one localized state into another, corresponding to the geometrical action. What one would really like to require was that the boost kept the state invariant; since that is impossible, the condition is relaxed to that of minimal deformation. In our approach, this axiom is not necessary, and the axiom of covariance under rotations is restored to its central position.

\end{document}